\documentclass[12pt]{article}

\usepackage[dvips]{graphics}

\def\beq   {\begin{equation}}
\def\eeq   {\end{equation}}
\def\beqd  {\begin{displaymath}}
\def\eeqd  {\end{displaymath}}
\def\beqaa {\begin{eqnarray}}
\def\eeqaa {\end{eqnarray}}

\def\ti  {\tilde}

\def\sq  {\ti q}
\def\st  {\ti t}
\def\sb  {\ti b}
\def\sg  {\ti g}

\def\nt  {\tilde\chi^0}
\def\ch  {\tilde\chi^\pm}
\def\chp {\tilde\chi^+}

\def\a   {\alpha}
\def\b   {\beta}
\def\t   {\theta}
\def\tst {\theta_{\st}}
\def\tsb {\theta_{\sb}}

\def\sz{\ifmmode{\tilde{\chi}^0} \else{$\tilde{\chi}^0$} \fi}
\def\sw{\ifmmode{\tilde{\chi}} \else{$\tilde{\chi}$} \fi}

\newcommand{\gsim}{\;\raisebox{-0.9ex}
           {$\textstyle\stackrel{\textstyle >}{\sim}$}\;}
\begin{document}
%------------------------------------------------------------------------
%\pagestyle{empty}

\vspace*{-1cm} 
\begin{flushright}
  TGU-27 \\
%  UWThPh-2000-46 \\
%  HEPHY-PUB 713/99 \\
%  AP-SU-99/01 \\
%  TU-561 \\
  hep-ph/0109083
\end{flushright}

%\vspace*{1.4cm}

\begin{center}

{\Large {\bf
Impact of bosonic decays on the search\\
for \boldmath{$\tilde t_1$} and \boldmath{$\tilde b_1$} squarks
}}\\

\vspace{10mm}

{\large 
K.~Hidaka}
\vspace{6mm}

\begin{tabular}{l}
{\it Department of Physics, Tokyo Gakugei University, Koganei,
Tokyo 184--8501, Japan}\\
\end{tabular}
{\it E-mail: hidaka@u-gakugei.ac.jp}
\end{center}

%\vspace{3cm}
%\vfill

\begin{abstract} 
We show that the bosonic decays of the lighter top and bottom 
squarks, i.e. 
$\st_1 \to \sb_1 + (H^+ \ \mbox{or} \ W^+)$ and
$\sb_1 \to \st_1 + (H^- \ \mbox{or} \ W^-)$, 
can be dominant in a wide range of the MSSM parameters. 
Compared to the fermionic decays, such as $\st_1 \to b + \chp_j$, 
these bosonic decays can have significantly different decay 
distributions. We also show that the effect of the supersymmetric 
QCD running of the quark and squark parameters 
on the $\st_1$ and $\sb_1$ decay branching ratios is quite dramatic.
These could have an important impact on the search for $\st_1$ 
and $\sb_1$ and the determination of the MSSM parameters 
at future colliders.\\
(Invited talk at The 9th International Conference on Supersymmetry 
and Unification of Fundamental Interactions (SUSY'01), 11-17 June 2001, 
Dubna, Russia; to be published in the Proceedings (World Scientific Pub.)) 
\end{abstract}

%\newpage
\pagestyle{plain}
%\setcounter{page}{2}

% --- introduction ---

%
We study the decays of the lighter top and bottom squarks 
(i.e. $\st_1$ and $\sb_1$) in the MSSM. 
They can decay into fermions, i.e. 
a quark plus a gluino ($\sg$), neutralino ($\nt_i$) or chargino ($\ch_j$). 
They can also decay into
bosons \cite{BMP}: 
\vspace{-1mm}
\beq
  \begin{array}{lcl}
    \st_1 \to \sb_1 + (H^+ \ \mbox{or} \ W^+)\,, &\hspace{3mm}&
    \sb_1 \to \st_1 + (H^- \ \mbox{or} \ W^-)\,.
  \end{array}
  \label{eq:bmodes}
\eeq
\vspace{-1mm}
In case the mass difference between $\st_1$ and $\sb_1$ is sufficiently 
large \cite{Raby}, the decays of Eq.~(\ref{eq:bmodes}) are possible. 
Here we extend the analysis of \cite{BMP}. 

% --- formulae and notation ---

The squark mass matrix in the basis $(\sq_L^{},\sq_R^{})$ with 
$\sq=\st$ or $\sb$ is given by \cite{BMP}
\begin{equation}
  {\cal M}^2_{\sq}= 
     \left( 
            \begin{array}{cc} 
                m_{\sq_L}^2 & a_q m_q \\
                a_q m_q     & m_{\sq_R}^2
            \end{array} 
     \right)       
  \label{eq:f}
\end{equation}
\begin{eqnarray}
  m_{\sq_L}^2 &=& M_{\ti Q}^2 
                  + m_Z^2\cos 2\beta\,(I_3^{q_L} - e_q\sin^2\t_W) 
                  + m_q^2 \label{eq:g} \\
  m_{\sq_R}^2 &=& M_{\{\ti U,\ti D\}}^2  
                  + m_Z^2 \cos 2\b\, e_q\, \sin^2\t_W + m_q^2 
                                              \label{eq:h}\\[2mm]
  a_q m_q     &=& \left\{ \begin{array}{l}
                     (A_t - \mu\cot\beta)\;m_t~~(\sq=\st)\\
                     (A_b - \mu\tan\beta)\,m_b~~(\sq=\sb) \, .
                          \end{array} \right. \label{eq:i}
\end{eqnarray}
We treat the soft SUSY-breaking parameters $M_{\ti Q,\ti U,\ti D}$ 
and $A_{t,b}$ as free ones 
since the ratios $M_{\ti U}/M_{\ti Q}$, $M_{\ti D}/M_{\ti Q}$ and 
$A_t/A_b$ are highly model-dependent.
By diagonalizing the matrix (\ref{eq:f}) one gets the mass eigenstate 
$\sq_1^{}=\sq_L^{}\cos\t_{\sq}+\sq_R^{}\sin\t_{\sq}$.\\
We take $M'=(5/3)\tan^2\t_W M$  and
$m_{\sg}=(\alpha_s(m_{\sg})/\alpha_2)M$ 
with $M$, $M'$ and $m_{\sg}$ 
being the SU(2), U(1) gaugino and gluino mass, respectively. 
We denote the mass of the CP-odd Higgs boson $A^0$ as $m_A$.
Full expressions of the widths of the squark decays are given in \cite{BMP}.

In case $M_{\ti Q,\ti U,\ti D}$ are relatively large 
in Eqs.(\ref{eq:f}-\ref{eq:i}), 
for $M_{\ti U} > M_{\ti Q} \gg M_{\ti D}$ 
($M_{\ti D} > M_{\ti Q} \gg M_{\ti U}$) we have $m_{\st_1} \gg m_{\sb_1}$ 
($m_{\sb_1} \gg m_{\st_1}$), which may allow the bosonic decays of 
Eq.(\ref{eq:bmodes}). 
We consider two patterns of the squark mass 
spectrum: $m_{\st_1} \gg m_{\sb_1}$ with ($\st_1$,$\sb_1$) $\sim$ 
($\st_L$,$\sb_R$) for $M_{\ti U} \gg M_{\ti Q} \gg M_{\ti D}$, and 
$m_{\sb_1} \gg m_{\st_1}$ with ($\st_1$,$\sb_1$) $\sim$ 
($\st_R$,$\sb_L$) for $M_{\ti D} \gg M_{\ti Q} \gg M_{\ti U}$. 
Thus the bosonic decays 
considered here are basically the decays of $\st_L$ into $\sb_R$ and 
$\sb_L$ into $\st_R$.

The leading terms of the squark couplings to $H^\pm$ are 
given by 
\vspace{-1mm}
\beqaa
G(\st_1 \sb_1 H^\pm) &\sim& 
               h_t(\mu\sin\b + A_t\cos\b)\sin\tst\cos\tsb  \nonumber \\
 & & {} + h_b(\mu\cos\b + A_b\sin\b)\cos\tst\sin\tsb. \label{eq:GH+} 
\eeqaa
\vspace{-1mm}
The Higgs bosons $H^\pm$ couple mainly to $\sq_L \sq_R^\prime$ 
combinations. These couplings are proportional to the Yukawa couplings 
$h_{t,b}$ and the squark mixing parameters $A_{t,b}$ and $\mu$ 
( Eq.(\ref{eq:GH+})). Hence the widths of the squark decays into $H^\pm$ 
may be large for large $A_{t,b}$ and $\mu$. 
In contrast, the gauge bosons $W^\pm$ couple only to $\sq_L \sq_L^\prime$, 
which results in suppression of the decays into $W^\pm$. 
However, this suppression is largely compensated by a large extra factor 
steming from the contribution of the longitudinally polarized W boson 
radiation ($\sq_1 \to \sq_1^\prime W_L^\pm$). 
Hence the widths of the squark decays into $W^\pm$ may be large for a 
sizable $\sq_L^\prime-\sq_R^\prime$ mixing term $a_{q^\prime} m_{q^\prime}$. 
On the other hand, the fermionic decays are not enhanced for large $A_{t,b}$ 
and $\mu$. Therefore the branching ratios of the bosonic decays of 
Eq.(\ref{eq:bmodes}) are expected to be large for large $A_{t,b}$ and $\mu$ 
if the gluino mode is kinematically forbidden. 
%As for the $\tan\b$ dependence, 
%we expect that the branching ratio $B(\st_1 \to \sb_1 + (H^+,W^+))$ increases 
%with increasing $\tan\b$ while $B(\sb_1 \to \st_1 + (H^-,W^-))$ is rather 
%insensitive to $\tan\b$, as explained in \cite{stop1decay}. 

The widths of the $\st_1$ and $\sb_1$ decays into $H^\pm$ receive very 
large SUSY-QCD corrections for large $\tan\b$ in the on-shell (OS) renormalization 
scheme \cite{sq_to_H_QCDcorr}, making the perturbative calculation unreliable. 
This problem can be solved by carefully defining the relevant tree-level 
couplings in terms of appropriate running parameters and on-shell squark 
mixing angles $\t_{\sq}$ \cite{improvedQCDcorr}. 
Following Ref.\cite{improvedQCDcorr}, we calculate the tree-level 
widths of the squark decays by using the corresponding tree-level 
couplings defined in terms of the SUSY-QCD running parameters $m_q(Q)$ and 
$A_q(Q)$ (with Q = (on-shell mass of the decaying squark $m_{\sq_1 OS}$)), 
and the on-shell squark mixing angles $\t_{\sq}$. 
We call the widths thus obtained as 'renormalization group (RG) 
improved tree-level widths'. 
Our input parameters are all on-shell ones except $A_b$ which is a 
running one, i.e. they are $M_t,M_b,M_{\ti Q}(\st),M_{\ti U},M_{\ti D},
A_b(Q = m_{\sq_1 OS}),A_t,\mu,\tan\b,m_A$, and M. 
$M_{\ti Q}(\sq)$ is the on-shell $M_{\ti Q}$ for the $\sq$ sector. 
The procedure for getting all necessary on-shell and SUSY-QCD running 
parameters is given in \cite{improvedQCDcorr}. For the calculation of the 
Standard Model running quark mass $m_q(Q)_{SM}$ from the two-loop RG equations 
we use the two-loop running $\a_s(Q)$ as in \cite{improvedQCDcorr}. 
We take $M_t$=175GeV and $M_b$=5GeV.
We choose $M_{\ti Q}(\st)$ = $\frac{3}{4} M_{\ti U}$ = $\frac{3}{2} M_{\ti D}$ 
($M_{\ti Q}(\st)$ = $\frac{3}{2} M_{\ti U}$ = $\frac{3}{4} M_{\ti D}$) 
for $\st_1$ ($\sb_1$) decays, and  $A_b(Q = m_{\sq_1 OS}) = A_t \equiv A$ 
for $\sq_1$ decay, for simplicity. 
Moreover, we fix M=400GeV (i.e. $m_{\sg}$=1065GeV) and $m_A$=150GeV. 
Thus we have $M_{\ti Q}(\st)$, A, $\mu$ and $\tan\b$ as free parameters. 
In the plots we impose the theoretical and experimental constraints 
\cite{stop1decay}.

%\newpage
%
%------------------------------------------------------------------------
% Figures
%------------------------------------------------------------------------
%
% Figure 1 --------------------------------------------------
%
\begin{figure}[!htb]
\begin{center}
%\vspace{20mm}
\hspace{-15mm}
\scalebox{0.6}[0.6]{\includegraphics{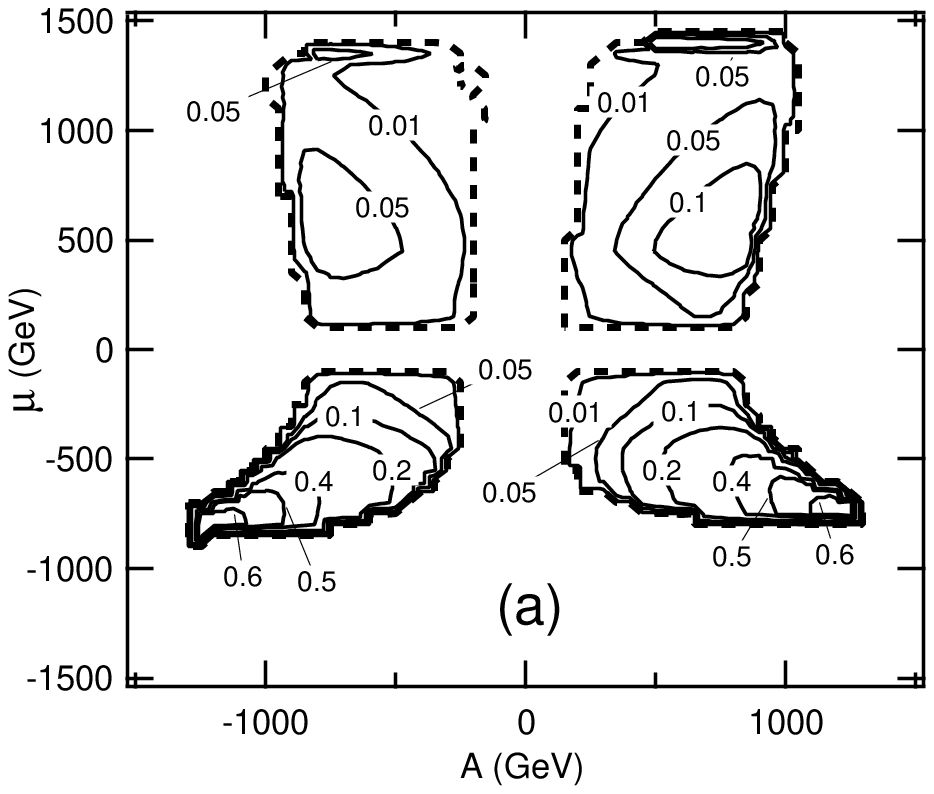}} 
\scalebox{0.6}[0.6]{\includegraphics{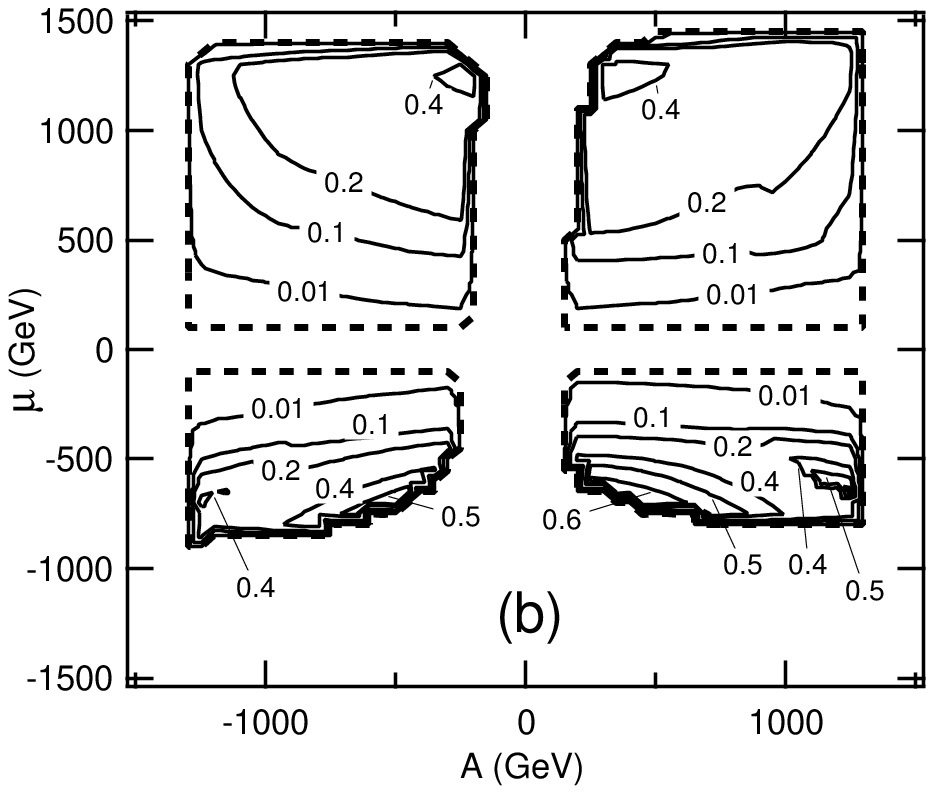}} \\
%\vspace{5mm}
\hspace{-15mm}
\scalebox{0.6}[0.6]{\includegraphics{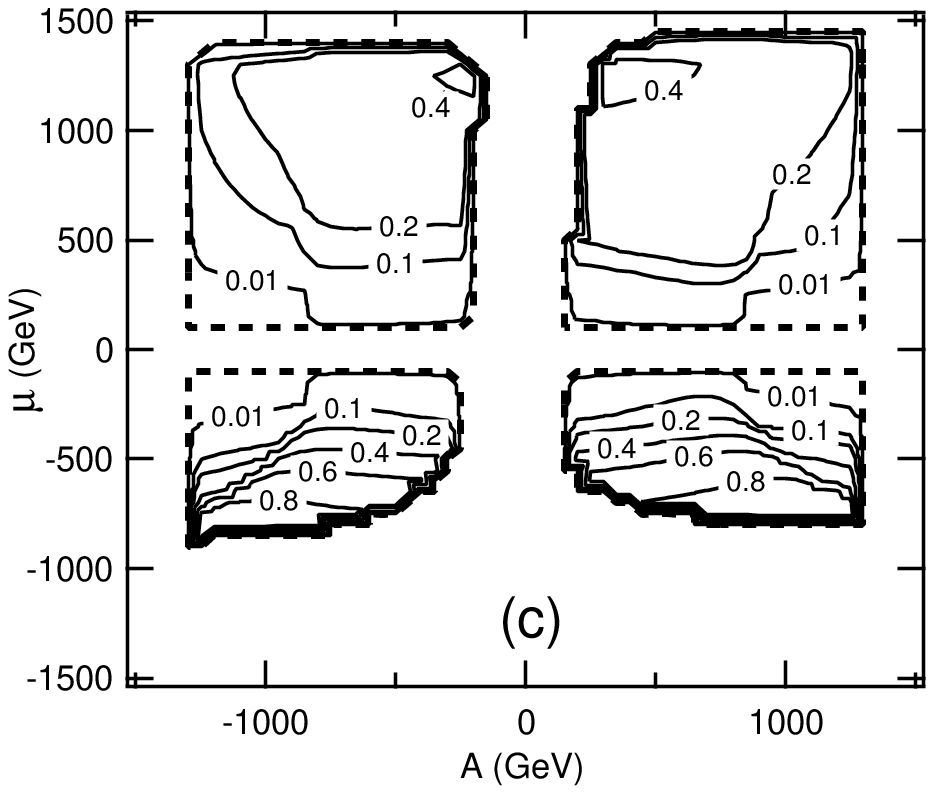}} 
\scalebox{0.6}[0.6]{\includegraphics{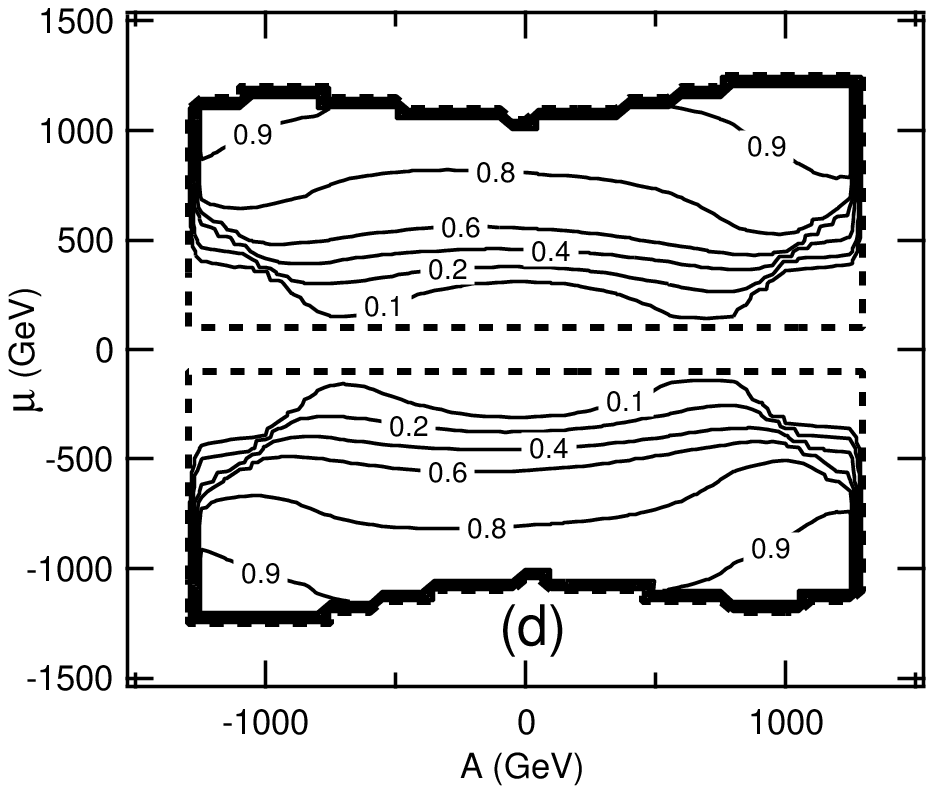}} \\
%\vspace{5mm}
%{\LARGE \bf Fig.1}
\end{center}

\caption{
Contours of the branching ratios at the RG-improved 
tree-level in the A--$\mu$ plane for $\tan\b=30$, and 
$M_{\ti Q}(\st)$ = $\frac{3}{4} M_{\ti U}$ = $\frac{3}{2} M_{\ti D}$ = 600GeV; 
(a)$B(\st_1 \to \sb_1 + H^+)$, (b)$B(\st_1 \to \sb_1 + W^+)$, and 
(c)$B(\st_1 \to \sb_1 + (H^+,W^+))$. The regions outside of the dashed loops 
are excluded by the kinematics and/or the constraints given in the text. 
Contours of the corresponding branching ratio 
$B(\st_1 \to \sb_1 + (H^+,W^+))$ at the naive tree-level are shown in Fig.d.
 \label{fig:contours}
}
\end{figure}
%----------------------------------------------------------------------

In Fig.\ref{fig:contours} we plot in the A-$\mu$ plane the contours of 
the $\st_1$ decay branching ratios of the Higgs boson mode , 
the gauge boson mode, and the total 
bosonic modes $B(\st_1 \to \sb_1 + (H^+,W^+)) \equiv 
B(\st_1 \to \sb_1 + H^+) + B(\st_1 \to \sb_1 + W^+)$  
at the RG-improved tree-level. 
We show also those of the corresponding branching ratio 
$B(\st_1 \to \sb_1 + (H^+,W^+))$ at the naive 
(unimproved) tree-level, where all input parameters are bare ones 
(see Eqs.(\ref{eq:f}-\ref{eq:i})). 
We see that the $\st_1$  decays into bosons are dominant in a large region of 
the A-$\mu$ plane, especially for large $|A|$ and/or $|\mu|$, as we 
expected. Comparing Fig.1.c with Fig.1.d we find that the effect 
of running of the quark and squark parameters 
$(m_q(Q),A_q(Q),M_{\ti Q,\ti U,\ti D}(Q))$ is quite dramatic. 
For $\sb_1$ decays we have obtained similar results to those for the $\st_1$ 
decays \cite{stop1decay}.

In Fig.3 of Ref. \cite{stop1decay} we show the individual branching ratios 
of the $\st_1$ and $\sb_1$ decays as a function of $\tan\b$ 
for $(A, \mu, M_{\ti Q}(\st))$=(-800, -700, 600)GeV and (800, 800, 600)GeV, 
respectively. We find that the branching ratios of the $\st_1$  decays 
into bosons increase with increasing $\tan\b$ and become dominant for 
large $\tan\b$ $(\gsim 20)$, while the $\sb_1$ decays into bosons are 
dominant in the entire range of $\tan\b$ shown, 
as expected \cite{stop1decay}. 

We find that the dominance of the bosonic modes is fairly insensitive 
to the choice of the values of $m_A$, M, and the ratio $A_b(Q)/A_t$. 

% --- conclusions ---
In conclusion, we have shown that the $\st_1$ and $\sb_1$ decays into Higgs or 
gauge bosons can be dominant in a fairly wide MSSM parameter region 
with large mass difference  between $\st_1$ and $\sb_1$, large $|A_{t,b}|$ 
and/or $|\mu|$, and large $m_{\sg}$ (and large $\tan\b$ for the $\st_1$ decay). 
%due to the large Yukawa couplings and mixings of $\st$ and $\sb$. 
%can be dominant in a significant portion of the MSSM parameter space 
%due to the large Yukawa couplings and mixings of $\st$ and $\sb$. 
Compared to the conventional fermionic decays these bosonic decays can have 
significantly different decay distributions \cite{stop1decay}. 
We have also shown that the effect of 
the SUSY-QCD running of the quark and squark parameters on the $\st_1$ and 
$\sb_1$ decays is quite dramatic. 
These could have an important impact 
on the searches for $\st_1$ and $\sb_1$ and 
on the determination of the MSSM parameters at future colliders.

\vspace{-3mm}
\section*{Acknowledgments}
\vspace{-3mm}
The author thanks A. Bartl very much for enjoyable 
collaborataion \cite{stop1decay}.


\vspace{-3mm}

\begin{thebibliography}{99}
\vspace{-3mm}
%1
\bibitem{BMP} 
A. Bartl, W. Majerotto, and W. Porod, Z. Phys. C 64 (1994) 499; 
C 68 (1995) 518 (E).

%2
\bibitem{Raby} 
For example, R. Dermisek, a talk in this Conference.

%3
\bibitem{stop1decay} 
K. Hidaka and A. Bartl, Phys. Lett. B501 (2001) 78; hep-ph/0012021.

%4
\bibitem{sq_to_H_QCDcorr}
A. Bartl, H. Eberl, K. Hidaka, S. Kraml, W. Majerotto, W. Porod, 
and Y. Yamada, Phys. Rev. D 59 (1999) 115007.
 
%5
\bibitem{improvedQCDcorr} 
H. Eberl, K. Hidaka, S. Kraml, W. Majerotto, 
and Y. Yamada, Phys. Rev. D62 (2000) 055006.

\end{thebibliography}
\end{document}